\documentclass[aps,pra,reprint,superscriptaddress,amsmath,showpacs,amsfonts,amssymb]{revtex4-1}

\usepackage{graphicx}
\usepackage{bm}
\usepackage{color}
\usepackage{MnSymbol}
\usepackage{enumerate}
\usepackage{bbold}

\def\ra{\rangle}
\def\cur{\mathcal{I}}
\def\be{\begin{equation}}
\def\ee{\end{equation}}

\newcommand{\bigjprob}{{\mathcal{P}}}

\newcommand{\op}[1]{\hat{ #1}}                

\newcommand{\Tr}[1]{\text{Tr}(#1)}               

\newcommand{\dd}{\mathrm{d}}

\newcommand{\drv}[1]{\frac{\delta}{\delta #1}}
\newcommand{\bbm}[1]{\bar{\bm #1}}

\begin{document}

\title{Action principle for continuous quantum measurement}

\author{A. Chantasri}
\affiliation{Department of Physics and Astronomy and Rochester Theory Center, University of Rochester, Rochester, New York 14627, USA}
\author{J. Dressel}
\affiliation{Department of Physics and Astronomy and Rochester Theory Center, University of Rochester, Rochester, New York 14627, USA}
\author{A. N. Jordan}
\affiliation{Department of Physics and Astronomy and Rochester Theory Center, University of Rochester, Rochester, New York 14627, USA}
\affiliation{Institute of Quantum Studies, Chapman University, 1 University Drive,
Orange, California 92866, USA}

\date{\today}

\begin{abstract}
We present a stochastic path integral formalism for continuous quantum measurement that enables the analysis of rare events using action methods. By doubling the quantum state space to a canonical phase space, we can write the joint probability density function of measurement outcomes and quantum state trajectories as a phase space path integral. Extremizing this action produces the most-likely paths with boundary conditions defined by preselected and postselected states as solutions to a set of ordinary differential equations. As an application, we analyze continuous qubit measurement in detail and examine the structure of a quantum jump in the Zeno measurement regime.
\end{abstract}


\maketitle

\section{Introduction}
There are qualitatively new features in fundamental quantum physics that appear in generalized (or weakened) measurements that are no longer simple projections \cite{BookNielsen,BookBreuer,BookWiseman}. For example, such measurements can be conditionally reversed \cite{Korotkov2006,*Korotkovexp2008}, they can approximately measure conjugate observables simultaneously \cite{Nazarov2008,Ruskov2010}, and they give a new perspective on the Heisenberg uncertainty relation \cite{Ozawa2003,*Erhart2012,*Rozema2012,*Weston2013}.  A sequence of weak measurements can also be made effectively continuous \cite{Mensky1979,*Mensky1994,Barchielli1982, *Diosi1988,Goetsch1994,Wiseman1996,Gurvitz1997,Levinson1997,Korotkov1999,*Korotkov2001,Jacobs2006,Nori2009,*Nori2009-2}, producing monitored state evolution in the form of a quantum stochastic process.  Importantly, such continuous monitoring opens the possibility of feedback control, where parameters in the system Hamiltonian are dynamically changed in response to the measurement record \cite{BookWiseman}.  This idea has been applied, for example, to rapid state purification \cite{Jacobs2003,Combes2006,Wiseman2006,Jordan2006,Combes2008,Wiseman2008,Ruskov2012}, as well as the stabilization of coherent oscillations \cite{Ruskov2002,Korotkov2005}, which was recently demonstrated with a superconducting transmon qubit \cite{Korotkov2012}.

For analogous classical stochastic processes, an important situation arises when the physics is sensitive to rare events starting and ending at certain points in phase space.  One well-known example is that of activation over a barrier, where the particle subjected to random forces begins at the bottom of the meta-stable well and ends at the saddle point, taking the particle out of the bound region \cite{Hanngi1990}.  Such rare events may be analyzed in the classical case by introducing a canonical phase space structure and minimizing the action subject to certain conditions (e.g. \cite{Dykman1994,Jordan2003,*Jordan2004,*JordanSuk2004,*SukhorJordan2007,Elgart2004,Bernard2006,Sinitsyn2009}). A similar situation arises in continuous quantum measurement when one is concerned with preparing a particular quantum state and then subsequently finding a rare final (postselected) state at a later time.  However, it has not been clear how to construct a quantum phase space that admits such an action principle suitable for studying these rare quantum events. 

The purpose of this article is to introduce and illustrate precisely such a phase space action principle over a doubled quantum state space.  This action principle is derived from a stochastic path integral formulation of the continuous measurement process that can admit additional boundary conditions, such as postselections.  This approach complements and reproduces the known formulation in terms of stochastic Schr\"odinger (or master) equations as a special case in the absence of postselection.  Notably, the phase space formulation enables the derivation of statistical quantities that may be difficult to calculate using the stochastic master equation approach, such as the average of all trajectories that satisfy a postselection condition.

Path integral approaches to continuous quantum measurement have a long history.  Mensky, for example, discussed a restricted Feynman path integral for continuous quantum measurement over three decades ago \cite{Mensky1979,*Mensky1994}.  Barchielli \emph{et al.} constructed a similar path integral, as did Caves \cite{Caves1986,*Caves1987}, which were subsequently related to the It\^o stochastic calculus by Di\'osi \cite{Barchielli1982,*Diosi1988}.  Wei and Nazarov recently discussed a different approach to continuous measurements using the Keldysh path integral technique \cite{Nazarov2008}.  Breuer and Petruccione discuss a related Hilbert space path integral formulation in their book \cite{BookBreuer}. However, our approach is qualitatively different from these previous works, since we consider the probability distribution function for paths of a (potentially mixed) quantum state through a canonically doubled state space.

Importantly, by doubling the state space to a canonical phase space, our extension permits extremizing the stochastic action to identify the most likely paths through a quantum phase space that satisfy both pre- and postselection boundary conditions.  These most likely paths are the solutions of ordinary (non-stochastic) differential equations, so can be illustrated as a phase space portrait and readily analyzed using classical methods.  As a paradigmatic demonstration, we consider the continuous measurement of a solid-state qubit and find its most probable dynamics in demolition and non-demolition regimes.  We also examine the structure of its rare quantum jump events in the Zeno measurement regime as particular paths through the phase space.

\section{Theoretical formulation}
\subsection{Stochastic path integral}
We consider a quantum system weakly coupled to a detector with discretized measurement readouts denoted by a set $\{r_k\}_{k=0}^{n-1}$. Each $r_k$ is assumed independent, and is obtained from the detector between time $t_k$ and $t_{k+1}=t_k+\delta t$. The quantum system state at these times is denoted by $\{ \op{\rho}_k\}_{k=0}^{n}$ where $\op{\rho}_0$ is an initial state and each $\op{\rho}_k$ is a state at time $t_k$ updated according to the prior measurement outcomes and the Hamiltonian evolution. We parametrize the density operator $\op{\rho}$ as a vector ${\bm q}$, where the components are the expansion coefficients in some orthogonal operator basis, such as the $d\equiv N^2-1$ generalized Gell-Mann matrices, ${\op \sigma}_j$ \cite{NLevelBloch} of a $N$ state system, $\op{\rho} =\frac{1}{N}(\hat{\mathbb{1}}+ \sum_{j=1}^d q_j {\op \sigma}_j)$, where $\hat{\mathbb{1}}$ is the identity. For a two-state system, the matrices $\op{\sigma}_j$ for $j=1,2,3$ are the Pauli matrices and $\bm{q}= (x,y,z)$ is a vector in Bloch sphere coordinates.

We are interested in applying initial and final boundary conditions to the parametrized quantum states $\bm{q}_0=\bm{q}_I$ and $\bm{q}_n=\bm{q}_F$.
The joint probability density function (PDF), $\bigjprob \equiv  P(\{ \bm{q}_k \},\{ r_k \},\bm{q}_F|\bm{q}_I)$, of all measurement outcomes $\{ r_k \}$, the quantum states $\{ \bm{q}_k \}$ and the chosen final state $\bm{q}_F$, conditioned on the initial state $\bm{q}_I$, is given by
\begin{align}\label{eq-mainpdf}
\bigjprob = \delta^d(\bm{q}_0-\bm{q}_I)\delta^d(\bm{q}_n-\bm{q}_F)
\prod_{k=0}^{n-1}P(\bm{q}_{k+1},r_k|\bm{q}_k).
\end{align}
Note that $d$ is the dimension of the vector $\bm{q}_k$. The conditional PDFs appearing in \eqref{eq-mainpdf} can be factored into products of two terms
\be
P(\bm{q}_{k+1},r_k|\bm{q}_k) = P(\bm{q}_{k+1}|\bm{q}_k,r_k)\, P(r_k|\bm{q}_k).
\label{twoterms}
\ee
The first term in \eqref{twoterms} describes the (deterministic) state update given the occurrence of result $r_k$, and is written as a $\delta$ function imposing the constraint $\bm{q}_{k+1} = \bm{q}_k +\delta t \bm{\mathcal{L}}[\bm{q}_k,r_k] + \mathcal{O}(\delta t^2)$, where $\bm{\mathcal{L}}$ is a vector functional describing the first order change in the state $\bm{q}_k$. We express these $\delta$ functions in Fourier form $\delta(q) = (1/2 \pi i) \int_{-i \infty}^{i \infty}  e^{-p q}\,\dd p$ with conjugate variables $\bm{p}_k$ integrated along contours with end-points at $\pm i \infty$. For example, we write $P(\bm{q}_{k+1} | \bm{q}_k, r_k)$ $= (1/2 \pi i)^{d}\int_{-i\infty}^{i\infty} \dd^d p_k \exp\big[-\bm{p}_k \cdot (\bm{q}_{k+1} -\bm{q}_k -\delta t \bm{\mathcal{L}}[\bm{q}_k,r_k] + \mathcal{O}(\delta t^2))\big]$. The second term in \eqref{twoterms} characterizes the probability of the measurement outcome $r_k$ given a quantum state $\bm{q}_k$, which we can also write in exponential form, $P(r_k|\bm{q}_k) \propto \exp\big[ \delta t\mathcal{F}[\bm{q}_k,r_k] + {\cal O}(\delta t^2)\big]$, where $\mathcal{F}[\bm{q}_k,r_k]$ is the linear order expansion of $\ln P(r_k|\bm{q}_k)$ in $\delta t$.

By taking the continuum limit $\delta t \to 0$, $n \to \infty$ and setting $t_0=0, t_n=T$, we obtain a stochastic path integral representation of the PDF \eqref{eq-mainpdf},
\begin{equation}\label{eq-pathint}
\bigjprob = \!\int \!\!\! {\cal D} \bm{p}\, e^{\cal S}=\!\int \!\!\!{\cal D} \bm{p}\,\exp \bigg[ \int_{0}^{T}\!\!\!\! \dd t \big(-  \bm{p}\cdot\dot{\bm{q}} +\mathcal{\mathcal{H}}[\bm{q},\bm{p},r]\big)\bigg],
\end{equation}
where the functional measure ${\cal D}\bm{p}$ absorbs the constant factor $\lim_{n \to \infty}(1/2 \pi i)^{d(n+2)}$ and other divergent constants. More details of the derivation of Eq.~\eqref{eq-pathint} using discretized variables are provided in Appendix \ref{appendderivation}. The stochastic action ${\cal S} = {\cal S}[\bm{q},\bm{p},r]$ is a functional of $\bm{q}(t),\bm{p}(t)$ and $r(t)$, as is the stochastic Hamiltonian
\begin{multline}\label{eq-H}
\mathcal{H}[\bm{q},\bm{p},r]= \bm{p}\cdot\bm{\mathcal{L}}[\bm{q},r]+ \mathcal{F}[\bm{q},r]+\\ -\bm{p}\cdot(\bm{q}-\bm{q}_I)\delta(t) - \bm{p}\cdot(\bm{q}-\bm{q}_F)\delta(t-T).
\end{multline}
The functions $\bm{q}(t)$ and $\bm{p}(t)$ act as effective coordinates and canonically conjugate momenta for the state space.

The expectation value of an arbitrary functional $\mathcal{A}[\bm{q},r]$ can now be computed from the PDF \eqref{eq-pathint} as a path integral $\int \! {\cal D}\bm{q} {\cal D}r {\cal A}[\bm{q},r] P(\bm{q}, r, \bm{q}_F | \bm{q}_I)$. For the case without postselection, we can set $\bm{p}(T)=0$ and compute unconditioned averages and correlation functions using diagrammatic perturbation expansions, which will be published at a later time. However, an expectation value conditioned on the pre- and postselected states can also be found from the conditioned PDF, $P(\bm{q},r|\bm{q}_I,\bm{q}_F)$, which is derived from the joint PDF \eqref{eq-pathint} using Bayes' rule.

\subsection{Most likely paths}
For chosen initial and final quantum states, we wish to determine the path $\bm{q}(t)$ and its corresponding measurement record $r(t)$ that give the maximal contribution to the integral $\int\! \mathcal{D}\bm{q} \mathcal{D}r \mathcal{D}\bm{p} \,e^{\mathcal{S}}$$= P(\bm{q}_F|\bm{q}_I)$. This path can be derived by extremizing the action ${\cal S}[\bm{q},\bm{p},r]$. Taking functional derivatives of the action and setting them to zero leads to the ordinary differential equations (ODEs)
\begin{subequations}\label{eq-sadpoint}
\begin{gather}
-\dot{\bm{q}} +\bm{\mathcal{L}}[\bm{q},r] =0, \\ 
 \dot{\bm{p}} + \drv{\bm{q}} \big(\bm{p}\cdot\bm{\mathcal{L}}[ \bm{q},r] \big)+ \drv{\bm{q}}\mathcal{F}[\bm{q},r]=0,\\ 
\drv{r}\big(\bm{p}\cdot\bm{\mathcal{L}}[\bm{q},r]\big)
+\drv{r}\mathcal{F}[\bm{q},r] =0,
\end{gather}
\end{subequations}
with the forced boundary conditions $\bm{q}(0)=\bm{q}_I$ and $\bm{q}(T) = \bm{q}_F$. Notably, both initial and final conditions can be imposed on $\bm{q}(t)$ in \eqref{eq-sadpoint} due to the additional integration constants from the canonical momenta $\bm{p}(t)$. The solution to \eqref{eq-sadpoint} gives the most-likely path, denoted by $\bbm{q},\bbm{p},\bar{r}$, for which $\mathcal{H}[\bbm{q},\bbm{p},\bar{r}]$ is a constant of motion. Note that the integration contours of $\int \! \mathcal{D}\bm{q} \mathcal{D}r \mathcal{D}\bm{p} \, e^{\mathcal{S}}$ can always be chosen to pass through these extremal points.

\section{Application to continuous qubit measurement}
We now apply this formalism to a solid-state detection setup: a single electron in a double quantum dot (DQD) where the electron location is weakly measured by a capacitively coupled quantum point contact (QPC) \cite{Gurvitz1997,Levinson1997,Korotkov1999,*Korotkov2001,Nori2009}. This setup can be easily extended to, e.g., a transmon qubit \cite{Korotkov2012}. The qubit states $|1\rangle$ and $|2\rangle$ correspond to the two dot locations. The density operator $\op{\rho}$ is a $2\times 2$ matrix. We choose the Bloch vector parametrization $\bm{q}=(x,y,z)$, using the Pauli matrices where $(0,0,1)$, $(0,0,-1)$ correspond to the states $|1\rangle, |2\rangle$, respectively. The Hamiltonian evolution of the qubit is determined by $\op{H}=(\epsilon/2) \op{\sigma}_3 + (-\Delta/2) \op{\sigma}_1$, where $\epsilon$ is an energy asymmetry and $\Delta$ is a tunneling strength.

The average current passing through the QPC, $\cur \equiv \cur(t)$, between time $t$ and $t+\delta t$ is assumed to be Gaussian with a mean $\bar{\cur}_{1,2}$ that depends on the state of the qubit $|1\rangle, |2\rangle$. We define a unitless readout $r=(\cur-\bar{\cur}_0)/\Delta \bar{\cur}$ for the QPC where $\bar{\cur}_0 = (\bar{\cur}_1+\bar{\cur}_2)/2$ and $\Delta\bar{\cur} = (\bar{\cur}_1-\bar{\cur}_2)/2$. The probability of the outcome $r$ is given by $P(r\,|\,\op{\rho})$$=\Tr{\op{\rho} \,\op{\cal M}_{\delta t}^\dagger \op{\cal M}_{\delta t}}$, where we define the measurement operator $\op{\cal M}_{\delta t}$ $= (\delta t / 2 \pi \tau)^{1/4} \exp\big[-\frac{\delta t}{4 \tau}(r-\op{\sigma}_3)^2\big]$.  Here $\tau$ $=S_0/2\Delta \bar{\cur}^2$ is the characteristic measurement time for the QPC that can be related to the sensitivity $\Delta \bar{\cur}$ and the QPC shot noise spectral density $S_0$.
Expanding $\ln P(r\,|\,\op{\rho}) \approx - (r^2-2 r z+1)\delta t/(2 \tau)+(1/2)\ln(\delta t/2 \pi \tau)+ {\cal O}(\delta t^2)$ determines the functional $\mathcal{F}[\bm{q},r]=-(r^2- 2 \,r\, z + 1)/(2\, \tau)$. The divergent constant proportional to $\ln(\delta t)$ is absorbed by the measure $\mathcal{D}\bm{p}$.

The vector functional $\bm{\mathcal{L}}[\bm{q},r]$, describing the first order change in a quantum state, which in this case is the qubit state, is derived from the state transformation equation,
\begin{align}\label{eq-statetransform}
\op{\rho}(t+\delta t) = \frac{ \op{\cal U}_{\delta t}\op{\rho}(t) \op{\cal U}_{\delta t}^{\dagger}}{\Tr{\op{\cal U}_{\delta t} \op{\rho}(t) \op{\cal U}_{\delta t}^{\dagger}}},
\end{align}
where $\op{\cal U}_{\delta t} \equiv e^{-\frac{i}{\hbar} \op{H} \delta t} \op{\cal M}_{\delta t}$ is a product of unitary evolution due to the qubit Hamiltonian $\op{H}$ and the measurement operator. By expanding \eqref{eq-statetransform} to first order in $\delta t$ and taking the continuum limit, we obtain a master equation,
\begin{align}
\partial_t\hat{{\rho}} = - \frac{i}{ \hbar}\big[ \op{H}, \op{\rho} \big] + \frac{r}{2 \tau}\{ \op{\sigma}_3 , \op{\rho}\} -  \frac{r}{\tau}\langle \op{\sigma}_3 \rangle \op{\rho},
\label{me}
\end{align}
where $[,]$, $\{,\}$, and $\langle \op{\sigma}_3 \rangle = \Tr{\op{\sigma}_3 \op{\rho}}$ are the commutator, the anti-commutator and an expectation value of $\op{\sigma}_3$, respectively. Expressing the right hand side of \eqref{me} in Bloch vector coordinates gives the vector functional $\bm{\mathcal{L}}[\bm{q},r]$.

Setting $\hbar = 1$, the action $\mathcal{S}$ and the stochastic Hamiltonian $\mathcal{H}$ of the PDF then take the form (see Appendix \ref{appendqdqpc} for more details of the derivation), 
\begin{subequations}\label{eq-actionxyz}
\begin{align}
\mathcal{S}&= \int_0^T\!\!\! \dd t \big( \! -  p_x \,\dot{x} -  p_y\,\dot{y} -  p_z\, \dot{z} + \mathcal{H}\big),\\ \nonumber
\mathcal{H} &= p_x(-\,\epsilon\, y-x\, z\, r/\tau)  + p_y(+\,\epsilon\, x+\Delta\, z-y\,z\,r/\tau) \\
&\label{eq-Hamil}
 + p_z(-\Delta\, y + (1-z^2)\,r/\tau) -(r^2- 2 \,r\, z + 1)/2 \tau,
\end{align}
\end{subequations}
where we have omitted the boundary condition terms.  Extremizing this action as in Eq.~\eqref{eq-sadpoint} produces the following $3+3$ ODEs and $1$ constraint,
\begin{subequations}\label{eq-ode1}
\begin{align}
\dot{x}&= -\, \epsilon\, y - x \,z \,r/\tau,\\
\dot{y}&=+\,\epsilon \,x +\Delta \,z - y \,z\, r/\tau, \\
\dot{z}&=-\,\Delta\, y +(1-z^2)\,r/\tau,\\
\dot{p}_x&=-\,\epsilon \, p_y + p_x\,z\, r/\tau, \\
\dot{p}_y&=+\,\epsilon  \,p_x+\Delta \, p_z +  p_y \,z\,  r/\tau, \\
\dot{p}_z&= -\,\Delta \, p_y + ( p_x \,x  +  p_y \,y  +2  \,p_z \,z -1)\,r/\tau,\\
r &=z +   p_z \,(1-z^2) -  p_x \,x \,z -  p_y\, y\, z,
\end{align}
\end{subequations}
with the (possibly mixed state) boundary conditions $\bm{q}_0 =(x_I,y_I,z_I)$ and $\bm{q}_F=(x_F,y_F,z_F)$.
\begin{figure}
\includegraphics[width=8.5cm]{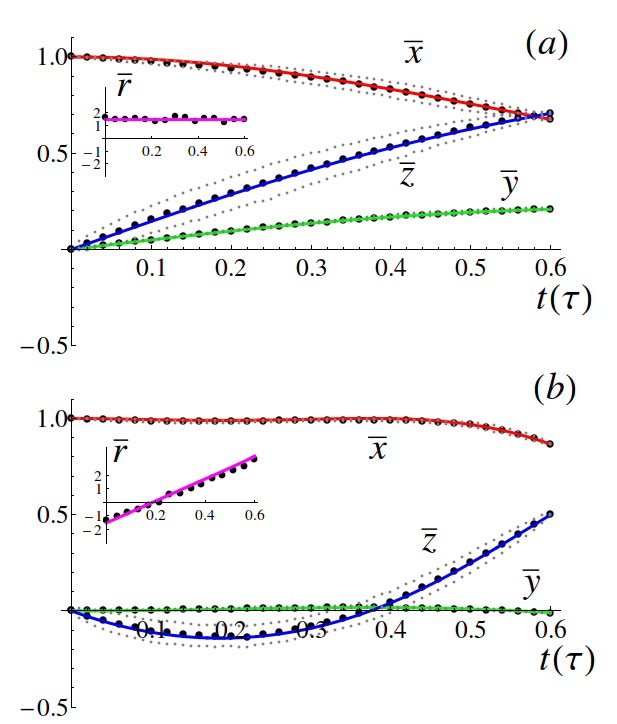}
\caption{\label{fig-xyz}(Color online) The numerical medians (black dots) of $x,y,z,r$ and the most-likely path from Eq.~\eqref{eq-ode1} (colored lines) $\bar{x},\bar{y},\bar{z},\bar{r}$ are shown for two scenarios: (a) $\Delta =0, \epsilon = 0.5 \,\tau^{-1}, \bm{q}_I=(1,0,0),\bm{q}_F\approx (0.7,0.2,0.7)$ (b) $\Delta = -\,0.5\, \tau^{-1},\epsilon =0, \bm{q}_I = (1,0,0),\bm{q}_F \approx (0.9,0,0.5)$. The postselected ensemble size  is $10^4$ with $\lambda = 0.02$, $\delta t =0.01 \,\tau$ and $T=0.6\,\tau$ (see text). The small dotted lines show the 40th and 60th percentile paths.}
\end{figure}

\textit{Quantum non-demolition (QND) measurement.---}
We can solve the differential equations \eqref{eq-ode1} analytically for the QND case when $\Delta =0$ \cite{BookBraginsky,BruneHaroche1994,JordanButt2005}. Eqs.~\eqref{eq-ode1} indicate that $\dot{r}$ vanishes in this case, so we conclude that its solution $r=\bar{r}$ is constant. With this insight, the most-likely path can then be solved immediately to give
\begin{subequations}\label{eq-minsol}
\begin{align}
\bar{x}(t) = \frac{x_I \,\cos \epsilon t - y_I\, \sin \epsilon t}{\cosh \bar{r} t/\tau + z_I\, \sinh \bar{r} t/\tau},\\
\bar{y}(t) =  \frac{y_I \,\cos \epsilon t + x_I\, \sin \epsilon t}{\cosh \bar{r} t/\tau + z_I \,\sinh \bar{r} t/\tau},\\
\bar{z}(t) =  \frac{z_I \,\cosh \bar{r} t/\tau + \sinh \bar{r} t/\tau}{\cosh \bar{r} t/\tau + z_I \,\sinh \bar{r} t/\tau},
\end{align}
\end{subequations}
where ${\bar r}$ can be found directly from the initial and final boundary conditions on the state, $\bar{r}$ $= \frac{\tau}{T} \tanh^{-1}\left(\frac{z_I-z_F}{z_I z_F-1}\right)$. We note that, in this QND case, the qubit coordinates can be solved directly from the boundary conditions, thus solutions of $p_x,p_y,p_z$ are not of particular interest and only presented in Appendix \ref{appendqnd}.

\textit{Numerical simulation.---}
To check the most-likely path, we numerically simulate qubit state trajectories using a Monte Carlo method. Starting with an initial state $\bm{q}_I=(x_I,y_I,z_I)$, a random outcome $r_0$ is drawn from a distribution $P(r_0|\bm{q}_I)$, and a new state $\bm{q}_1$ is computed from the state update equation \eqref{eq-statetransform}. Repeating this computation from $t_0=0$ to $t_{n}=T$ with time step $\delta t$ produces a single stochastic trajectory for $\bm{q}$. We postselect the ensemble of trajectories that conforms to the final boundary condition using the requirement $|\bm{q}_n-\bm{q}_F| \le \lambda$, where $\lambda$ is a postselection tolerance. The most-likely path is estimated from the statistical median of $\bm{q}$ at each time step. We use the median since it is more numerically robust than the mode. Fig.~\ref{fig-xyz} shows that these simulated medians agree quite well with solutions of the ODEs in \eqref{eq-ode1}.

\section{Anatomy of a quantum jump}
To further illustrate the action formalism, we consider the physics of the quantum Zeno effect:  an evolving qubit that is repeatedly measured on a time scale faster than the inverse Rabi oscillation frequency will be frozen in a particular state, only occasionally making a quantum jump to the orthogonal state on a longer time scale (e.g. \cite{Sudashan1977,*Kwiat1998}).  Here we analyze this situation in the continuous measurement case, restricting our attention to pure states for simplicity.  We set $\epsilon = 0$, $\tau \ll \Delta^{-1}$, which corresponds to the limit where the qubit measurement rate is much faster than the Hamiltonian dynamics.  

Considering pure states on the Bloch sphere, we re-parametrize the state as $x=0, y = \sin \theta, z = \cos \theta$, so $\theta = 0$ corresponds to state $|1\ra$ and $\theta = \pi$ corresponds to state $|2\ra$. Since the quantum state is now characterized by one variable, $\theta$, the master equation is quite simple, ${\dot \theta} = \Delta + \sin \theta\,  r /\tau$, and the action only depends on $\theta,p_\theta$ and $r$,
\begin{subequations}\label{eq-zenoaction}
\begin{align}
\mathcal{S}&= \int_0^T\!\! \!\dd t \,  \big(\! - p_\theta \,{\dot \theta} + {\cal H}\big),  \\
{\cal H} &=  p_\theta (\Delta -r \sin \theta /\tau) - (r^2 - 2 \,r \cos \theta +1)/2\tau.
\end{align}
\end{subequations}
Extremizing the action leads to two ODEs and a constraint,
\begin{subequations}\label{eq-odezeno}
\begin{align}
\dot{\theta} &= \Delta - \sin \theta \, r /\tau, \\
{\dot p}_\theta &= p_\theta\, r \cos \theta/\tau + r \sin \theta/\tau, \\
r &= \cos \theta - p_\theta \sin \theta.
\end{align}
\end{subequations}
After eliminating the $r$ variable, we have a Hamiltonian that is quadratic in the conjugate variable,  ${\cal H} = a\, p_\theta^2 + b\, p_\theta + c$, where $a = \sin^2 \theta/2 \tau$, $b = \Delta - \sin \theta \cos \theta /\tau$, $c = -\sin^2 \theta/2 \tau$. Since ${\cal H}$ is a constant of motion for the most probable path \eqref{eq-odezeno}, we can parametrize $p_\theta(\theta,E)$ as a function of $\theta$ and $E \equiv {\cal H}$.

\begin{figure}
\includegraphics[width=8cm]{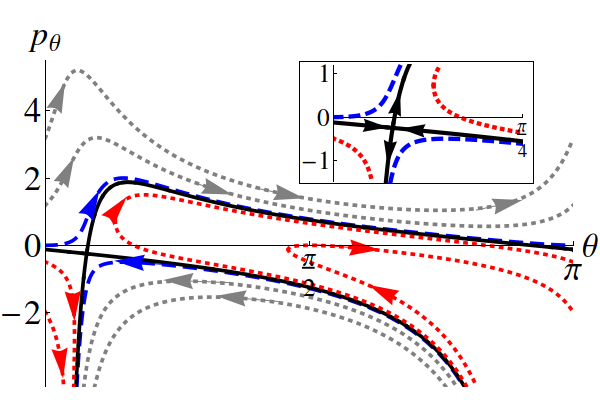}
\caption{\label{fig-phasespace}(Color online) The doubled state space phase portrait for different constant stochastic energies: $E<-\Delta^2 \tau/2$ (red dotted), $E=-\Delta^2 \tau/2$ (black solid), $E=0$ (blue dashed) and $E>0$ (gray dotted), for $\Delta =0.2 \,\tau^{-1}$. The dynamics is invariant under $\theta \to \theta + \pi$.}
\end{figure}

A phase space portrait $(p_\theta,\theta)$ of the dynamics is shown in Fig.~\ref{fig-phasespace}. There, the global dynamical structure can be seen for different constants of motion $E$, which we refer to as stochastic energy. We find that there is a critical value of the stochastic energy $E_c=-\Delta^2 \tau/2$ separating two types of paths: ones that can cross from $0$ to $\pi$ $(E>E_c)$, and ones that turn to the poles of origin $(E<E_c)$ (see inset of Fig.~\ref{fig-phasespace}). Note that the total time associated with a path from an initial $\theta_i$ to a final $\theta_f$ is $T=\int_{\theta_i}^{\theta_f} \dot{\theta}(\theta,E)^{-1}\dd\theta$ and the associated action is ${\cal S}=-\int_{\theta_i}^{\theta_f}\!p_\theta(\theta,E)  \dd\theta   + E T$. More details of the state dynamics on the phase space for different values of the stochastic energy are provided in Appendix \ref{quantumjump} and Supplementary Material \cite{Note1}.

Among all the constant energy curves in the phase portrait,  the one associated with $\mathcal{H}=E=0$ has the maximum probability since it minimizes the action $\mathcal{S}$ with respect to $\mathcal{H}$. This zero energy (instanton) solution $p_\theta^{in}$ is shown in Fig.~\ref{fig-phasespace} (blue dashed) and can be simplified in the limit $\Delta \tau \ll 1$,
\be
p_{\theta}^{in} \approx \begin{cases} 0, & \, 0 < \theta < \Delta \tau\\
 2 (\theta - \Delta \tau)/\theta^2, & \Delta \tau < \theta < \pi.
\end{cases}
\label{instantonline}
\ee
Since the action is the integral of the instanton line \eqref{instantonline}, we see that the state can transition from $\theta=0$ to $\theta = \Delta \tau$ with nearly unit probability, but then encounters a statistical barrier that requires a large fluctuation to overcome.  This can be interpreted as the state evolving unitarily to $\theta(t) = \Delta t$, but then collapsing back onto the $\theta=0$ point after one measurement time $\tau$.   Occasionally, the state can flip over to the opposite pole with a small rate. The area under the curve \eqref{instantonline} (the blue dashed curve in Fig.~\ref{fig-phasespace}) gives the action of this path, $\mathcal{S}_{in} \approx  2 \ln (\Delta \tau)$ according to \eqref{eq-zenoaction}, which is logarithmically divergent for small $\Delta$. In this limit, the time of this path is given by $T \approx 4 \tau \ln( 1/\Delta \tau)$.  This is a characteristic time taken by a quantum jump.  The instanton technique for finding the switching rate resembles standard tunneling rate calculations \cite{BookKleinert}, so the switching rate has the form $\gamma = \omega_{\text{att}} e^{\mathcal{S}_{in}}$, where $\omega_{\text{att}}$ is the attempt frequency. In this case, $\omega_{\text{att}}$ is the measurement rate, $\tau^{-1}$, so the switching rate is simply $\gamma = \Delta^2 \tau$.

\section{Conclusion}
We have developed a phase space stochastic path integral formalism for continuous
quantum measurement that enables the analysis of rare events using action techniques. As an example, we derived a set of ODEs that describe the most likely path the quantum state can take through state space between initial and final boundary conditions via an action principle. These equations describe a `middle way' between the well-known master equation for open quantum systems and stochastic master equations.  Our ODEs double the complexity of the standard master equation, but are much simpler to numerically integrate than the simulation of a large ensemble of stochastic trajectories, followed by postselection on a small fraction of realizations.  These ODE solutions give deep insight into the conditional dynamics of the measured quantum system, and also provide the most likely detector output corresponding to the imposed boundary conditions. 

Master equations are perhaps the most used tool in current research on quantum systems.  Our reformulation of the physics of continuously measured quantum systems in terms of the stochastic path integral \eqref{eq-pathint}-\eqref{eq-H} gives another way to compute averages and correlations functions of the detector output and the quantum state, as well as permitting the calculation of conditional quantities.  The assumptions of this method are quite general, and it may consequently be applied to a wide variety of physical systems.

\begin{acknowledgments}
This work was supported by National Science Foundation under Grant No. DMR-0844899 and by Development and Promotion of Science and Technology Talents Project Thailand.
\end{acknowledgments}

\appendix
\section{Derivation of the Stochastic Path Integral} \label{appendderivation}
Here we present the technical details of the stochastic path integral derivation. The joint PDF, $\bigjprob = P(\{ \bm{q_k}\},\{r_k\}, \bm{q}_F | \bm{q}_I)$, is a product of many different terms as shown in Eq.~\eqref{eq-mainpdf} and Eq.~\eqref{twoterms}. We first write the PDF for the deterministic state update from a quantum state $\bm{q}_k$ to $\bm{q}_{k+1}$ for $k=0,...,n-1$,
\begin{align}
\nonumber P(&\bm{q}_{k+1} | \bm{q}_k, r_k)  \equiv \delta^d (\bm{q}_{k+1} - \bm{\mathcal{E}}[\bm{q}_k,r_k]),\\
\nonumber &= (1/2 \pi i)^{d}\int_{-i\infty}^{i\infty} \dd^d p_k \exp\big[-\bm{p}_k \cdot (\bm{q}_{k+1} - \bm{\mathcal{E}}[\bm{q}_k,r_k])\big],
\end{align}
where $\bm{\mathcal{E}}[\bm{q}_k,r_k] \equiv \bm{q}_k + \delta t \bm{\mathcal{L}}[\bm{q}_k,r_k]+ \mathcal{O}(\delta t^2)$ and use $\dd^d p_k$ as an integration measure of the $d$-dimensional vector $\bm{p}_k$. We also rewrite the function $P(r_k |\bm{q}_k)$ $=\exp\big[\ln P(r_k |\bm{q}_k)\big]=\exp\big[\delta t \mathcal{F}[\bm{q}_k,r_k] +\,\, \cdots\, \big]$ in exponential form. The other two $\delta$ functions for the boundary conditions in Eq.~\eqref{eq-mainpdf} are also written in Fourier form,
\begin{align}
\nonumber \delta^d(\bm{q}_0-\bm{q}_I) &= (1/2 \pi i)^{d}\int_{-i\infty}^{i\infty} \dd^d p_{-1} \exp\big[-\bm{p}_{-1}\cdot (\bm{q}_0-\bm{q}_I)\big],\\
\nonumber \delta^d(\bm{q}_n-\bm{q}_F) &=(1/2 \pi i)^{d} \int_{-i\infty}^{i\infty} \dd^d p_{n} \exp\big[- \bm{p}_{n}\cdot (\bm{q}_n-\bm{q}_F)\big].
\end{align}
As a result, the PDF Eq.~\eqref{eq-mainpdf} is a product of all terms presented above,
\begin{align}\label{eq-discspi}
\bigjprob =C \idotsint\limits_{-i\infty}^{i\infty} \prod_{j=-1}^{n}\mathrm{d}^d p_j \exp\bigg[B +\sum_{k=0}^{n-1} (L_k+F_k) \bigg],
\end{align}
where,
\begin{equation}\nonumber
\begin{split}
B &= - \bm{p}_{-1} \cdot (\bm{q}_0-\bm{q}_I) - \bm{p}_n \cdot (\bm{q}_n-\bm{q}_F),\\
L_k & =  - \bm{p}_k \cdot (\bm{q}_{k+1}-\bm{\mathcal{E}}[\bm{q}_k,r_k]),\\
F_k &=\ln P(r_k | \bm{q}_k), \\
C & =(1/2 \pi i)^{d(n+2)},
\end{split}
\end{equation}
The stochastic path integral Eq.~\eqref{eq-pathint} is a continuous version of \eqref{eq-discspi} derived by taking the continuum limit $\delta t \rightarrow 0$ and $n \to \infty$. Neglecting second and higher order terms in $\delta t$ yields the replacements: $\bm{q}_{k+1}-\bm{\mathcal{E}}[\bm{q}_k,r_k] \approx \dot{\bm{q}}\dd t -  \bm{\mathcal{L}}[\bm{q},r]\dd t$ and $\ln P(r_k | \bm{q}_k) \approx \dd t \mathcal{F}[\bm{q},r]$. In general, there may be a ($r_k,\bm{q}_k$)-independent correction to the approximation of $\ln P(r_k | \bm{q}_k)$ that does not vanish in the continuum limit, but it can always be absorbed by the constant $C$. Note that we used the continuous coordinate function $\bm{q} = \bm{q}(t) = \lim_{\delta t \rightarrow 0}\{ \bm{q}_k \}$, conjugate function $\bm{p}=\bm{p}(t) = \lim_{\delta t \rightarrow 0}\{ \bm{p}_k \}$ and measurement record function $r=r(t) = \lim_{\delta t \rightarrow 0}\{ r_k \}$. The functional measure used in Eq.~\eqref{eq-mainpdf} is defined as $\mathcal{D}\bm{p} \equiv  \lim_{\delta t \to 0}C \prod_{j=-1}^{n}\mathrm{d}^d p_j$.

The most-likely path in Eq.~\eqref{eq-sadpoint} can be obtained by extremizing the exponent of \eqref{eq-discspi} over all variables $(k=0,...,n-1)$,
\begin{subequations}\label{eq-discode}
\begin{gather}
 -\bm{q}_{k+1} +\bm{\mathcal{E}}[\bm{q}_k,r_k]=0,\\
 -\bm{p}_{k-1}+ \frac{\partial}{\partial \bm{q}_k} \big(\bm{p}_k\cdot \bm{\mathcal{E}}[\bm{q}_k,r_k]\big)+\frac{\partial}{\partial \bm{q}_k}\big(\ln P(r_k | \bm{q}_k)\big)=0, \\
 \frac{\partial}{\partial r_k}\big(\bm{p}_k \cdot \bm{\mathcal{E}}[\bm{q}_k,r_k]\big) + \frac{\partial}{\partial r_k}\big(\ln P(r_k | \bm{q}_k)\big)=0,
\end{gather}
\end{subequations}
including the boundary conditions: $\bm{q}_0 =\bm{q}_I$, $\bm{q}_n = \bm{q}_F$. After taking the continuum limit, the Eqs.~\eqref{eq-discode} reduce to Eqs.~\eqref{eq-sadpoint}.

Since the solutions of \eqref{eq-discode} as well as \eqref{eq-sadpoint} extremize the action $\mathcal{S}$, a Taylor series expansion of the action up to second order around the extremal point $\bbm{q},\bbm{p},\bar{r}$, gives the integral $\int \mathcal{D}\bm{q} \mathcal{D}r \mathcal{D}\bm{p} \, e^{\mathcal{S}}$$= P(\bm{q}_F|\bm{q}_I)$ in the form,
\begin{align}\label{eq-integratep}
\nonumber P(\bm{q}_F|\bm{q}_I)& = e^{\mathcal{S}[\bbm{q},\bbm{p},\bar{r}]}\times\\
 &\int\!\! \dd \bm{\eta}\, \,e^{\frac{1}{2!}\,\bm{\eta}^\top\! \cdot \, \bm{D}^2 \mathcal{S}[\bbm{q},\bbm{p},\bar{r}]\,\cdot \,\bm{\eta}\, + \,\mathcal{O}(\bm{\eta}^3)},
\end{align} 
where the components of the vector $\bm{\eta}$ are all components of $(\bm{q}-\bbm{q})$, $(\bm{p}-\bbm{p})$ and $(r-\bar{r})$, and $\bm{D}^2 \mathcal{S}[\bbm{q},\bbm{p},\bar{r}]$ is a matrix of second-order partial derivatives of the action $\mathcal{S}$ evaluated at the extrema $\bbm{q},\bbm{p},\bar{r}$. In the saddle point method, one neglects the higher order terms keeping only second order contributions. However, here we focus on the leading term $e^{\mathcal{S}[\bbm{q},\bbm{p},\bar{r}]}$, which is the contribution from the most-likely path $\bbm{q},\bbm{p},\bar{r}$ to the probability $P(\bm{q}_F|\bm{q}_I)$, similar to the large deviation function studied in the macroscopic fluctuation theory \cite{Bernard2007}.

It is worth noting that the set of equations \eqref{eq-discode} can also be obtained through constrained optimization of the PDF directly, 
\begin{align}\nonumber
P(\{ r_k \} |\{\bm{q}_k\})& = \prod_{k=0}^{n-1} P(r_k |\bm{q}_k)=\exp\bigg[\sum_{k=0}^{n-1} \ln P(r_k|\bm{q}_k) \bigg],
\end{align}
where the conjugate variables $ \{\bm{p}_k \}$ play the role of Lagrange multipliers for the constraints $\bm{q}_0=\bm{q}_I$, $\bm{q}_{k+1} = \bm{\mathcal{E}}[\bm{q}_k,r_k]$ (for $k=0,...,n-1$), and $\bm{q}_n=\bm{q}_F$, respectively. Therefore, the maximum probability conforming to the constraints is,
\begin{align}
\text{Max}\big[P(\{ r_k \} |\{\bm{q}_k\})\big] = \exp\bigg[\sum_{k=0}^{n-1}  \ln P(\bar{r}_k|\bbm{q}_k) \bigg],
\end{align}
where $\bbm{q}_k$ and $\bar{r}_k$ are solutions of Eq.~\eqref{eq-discode}. This maximum probability is proportional to $e^{\mathcal{S}[\bbm{q},\bbm{p},\bar{r}]}$ in \eqref{eq-integratep} in the continuum limit.

\section{The continuous qubit measurement} \label{appendqdqpc}
We now show more details about the derivations of the action in Bloch coordinates Eq.~\eqref{eq-actionxyz}. In discrete form, the unitless measurement readout is given by $r_k= (\cur_k-\bar{\cur}_0)/\Delta \bar{\cur}$ where $\cur_k$ is an average current passing through the QPC between time $t_k$ and $t_{k+1}$. The probability density function is explicitly expressed in Bloch sphere coordinates $\bm{q}_k = (x_k,y_k,z_k)$ at time $t_k$
\begin{align}\label{eq-probr}\nonumber
P(r_k | \bm{q}_k)\, =\,\,& \sqrt{\frac{\delta t}{2 \pi \tau}}e^{-\frac{\delta t}{2 \tau}(r_k-1)^2}(1+z_k)/2 \\ \nonumber
&+\sqrt{\frac{\delta t}{2 \pi \tau}} e^{-\frac{\delta t}{2 \tau}(r_k+1)^2}(1-z_k)/2,\\ \nonumber
\ln P(r_k | \bm{q}_k) \approx \,   &  -\frac{\delta t}{2 \tau}  \big( r_k^2 - 2 r_k z_k + 1)
+ \frac{1}{2}\ln \left(\frac{\delta t}{2 \pi \tau}\right) + \mathcal{O}(\delta t^2) ,
\end{align}
where we find that ${\cal F}[\bm{q}_k,r_k] = - (r_k^2-2 r_k z_k +1)/(2 \tau)$. The constant term proportional to $\ln(\delta t)$ will be absorbed by the integration measure.

We take the state update Eq.~\eqref{eq-statetransform} in discrete form, express it in Bloch sphere coordinates, and take the first order expansion in $\delta t$,
\begin{subequations}
\begin{align}
x_{k+1} &=x_k+( - \epsilon y_k - x_k z_k r_k/\tau)\delta t,\\
y_{k+1} &=y_k +(+\epsilon x_k + \Delta z_k - y_k z_k r_k /\tau)\delta t, \\
z_{k+1} &=z_k+( - \Delta y_k + (1-z_k^2)r_k/\tau)\delta t,
\end{align}
\end{subequations}
where the right hand side is the vector function $\bm{\mathcal{E}}[\bm{q}_k,r_k]$. Substituting $\bm{\mathcal{E}}[\bm{q}_k,r_k]$ and ${\cal F}[\bm{q}_k,r_k]$ into \eqref{eq-discspi}, we obtain the action in the discrete form,
\begin{align}\label{eq-discaction}\nonumber
\mathcal{S} = B &+ \sum_{k=0}^{n-1}\bigg( -p_k^x\big(x_{k+1}-x_k -(-\epsilon y_k - x_k z_k r_k/\tau)\delta t\big) \\ \nonumber
&-p_k^y \big(y_{k+1}-y_k - (+\epsilon x_k + \Delta z_k - y_k z_k r_k /\tau)\delta t\big)\\ \nonumber
& -  p_k^z\big(z_{k+1}-z_k-(-\Delta y_k+ (1-z_k^2) r_k /\tau)\delta t\big)\\
&-(r_k^2-2 r_k z_k + 1)\delta t/2 \tau \bigg),
\end{align}
where $p_k^x,p_k^y,p_k^z$ are conjugate variables and $B$ is a term describing the boundary conditions $\bm{q}_0 = (x_I, y_I, z_I)$ and $\bm{q}_n = (x_F,y_F,z_F)$ as defined in \eqref{eq-discspi}.

\section{The QND measurement case}\label{appendqnd}
In the QND case when $\Delta =0$, we can solve for the analytic solutions $\bar{x},\bar{y},\bar{z}$ as in Eqs.~\eqref{eq-minsol} and solve for solutions of conjugate variables $\bar{p}_x,\bar{p}_y,\bar{p}_z$
\begin{subequations}
\begin{align}
\bar{p}_x &=  \frac{p_{xI} \,\cos \epsilon t - p_{yI}\, \sin \epsilon t}{(\cosh \bar{r} t/\tau + z_I\, \sinh \bar{r} t/\tau)^{-1}},\\
\bar{p}_y &= \frac{p_{yI} \,\cos \epsilon t +p_{xI}\, \sin \epsilon t}{(\cosh \bar{r} t/\tau + z_I\, \sinh \bar{r} t/\tau)^{-1}},\\
\bar{p}_z &= \frac{\bar{r}-\bar{z}+\bar{p}_x\,\bar{x}\, \bar{z} +\bar{p}_y\, \bar{y}\,\bar{z}}{1-\bar{z}^2},
\end{align}
\end{subequations}
where $p_{xI}, p_{yI},p_{zI}$ are arbitrary constants. Generally the conjugate variables $p_x, p_y, p_z$ are constrained by the choice of final boundary conditions $x_F, y_F, z_F$ for the coordinates, so the six coupled differential equations must be solved simultaneously.  However, in the special case of $\Delta = 0$ then the most likely readout $\bar{r}$ is a constant, so the constraint becomes infinitely degenerate.  As a result, solutions for $p_x, p_y, p_z$ can be obtained through direct integration, which produces arbitrary integration constants $p_{xI}, p_{yI}, p_{zI}$ that indicate the degeneracy.

\begin{figure}
\includegraphics[width=8cm]{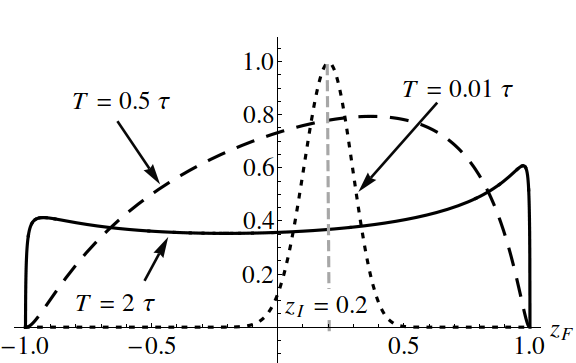}
\caption{This shows the exponential of the extremized action Eq.~\eqref{eq-logprobqnd} as a function of $z_F$ for the QND case  when $z_I = 0.2$  plotting at three different time $T = 0.01 \,\tau$ (dotted) $T=0.5 \,\tau$ (dashed) and $T=2 \,\tau$ (solid).}
\label{fig-prob}
\end{figure}

By substituting the extremum $\bar{x},\bar{y},\bar{z},\bar{r}$ back to the action in Eqs.~\eqref{eq-actionxyz}, all terms except the last term in the stochastic Hamiltonian vanish and we obtain the optimized action,
\begin{equation}\label{eq-logprobqnd}
\begin{split}
\mathcal{S}[\bar{x},\bar{y},\bar{z},\bar{r}]= -\frac{T}{2 \tau}(\bar{r}^2+1) +\frac{1}{2} \ln \bigg(\frac{1-z_I^2}{1-z_F^2}\bigg),
\end{split}
\end{equation}
where $\bar{r} = \frac{\tau}{T} \tanh^{-1}\big(\frac{z_I-z_F}{z_I z_F-1}\big)$. Here, we can see how the leading term of $P(\bm{q}_F|\bm{q}_I)$ in Eq.~\eqref{eq-integratep} changes as we vary the boundary conditions $\bm{q}_F$, $\bm{q}_I$. We plot the exponential of the extremized action \eqref{eq-logprobqnd} as a function of $z_F$ setting $z_I = 0.2$ in Fig.~\ref{fig-prob}. We see that at very short time, $T=0.01 \,\tau$, the final state $z_F$ is still mostly around the initial state. As the time grows to $T=0.5 \,\tau$, $T=2  \,\tau$, the curves gets broader and the most probable final states move toward either of the poles $z_F = \pm 1$. The asymmetry of the long time distribution is due to the initial state $z_I \ne 0$.

\section{Quantum jump analysis} \label{quantumjump}
Further insight can be gained about the critical stochastic energy $E_c = -\Delta^2 \tau/2$ mentioned in the main text by extremizing the equations of motion.  There exists a local stationary point (see inset of Fig.~\ref{fig-phasespace} where the nullclines of the dynamics cross, ${\dot \theta} =0$, ${\dot p}_\theta =0$.  This condition gives stationary values for the fixed point $\theta_s, r_s,p_{\theta,s}$, found from the Eqs.~\eqref{eq-odezeno},
\begin{align}
\nonumber \Delta \tau &=  r_s \sin \theta_s, \\
\nonumber 0 &= p_{\theta,s}\,r_s \cos \theta_s + r_s \sin \theta_s.
\end{align}
Together with the last condition in Eq.~\eqref{eq-odezeno}, $r_s= \cos \theta_s -  p_{\theta,s} \sin \theta_s$, we can solve for the three unknowns,
\begin{subequations}\label{values}
\begin{align}
\theta_s &= \tan^{-1} \Delta \tau, \\
p_{\theta,s} &= - \Delta \tau, \\
r_s &= \sqrt{1 + \Delta^2 \tau^2}.
\end{align}
\end{subequations}
The leading term in the action associated with this stationary point is 
\begin{subequations}
\begin{align}
\mathcal{S}[\theta_s, r_s]&= -\int_{t_i}^{t_f}\!\!\! \dd t \, \frac{1}{2 \tau}\big(r_s^2 - r_s \cos \theta_s + 1\big) \\
&= - \frac{\Delta^2 \tau }{2}(t_f-t_i),
\end{align}
\end{subequations}
assuming that the qubit stays at the steady state from $t_i$ to $t_f$.

This approach gives us another way to understand the nature of the critical stochastic energy $E_c $. For the particular value of the detector output $r_s$, the Hamiltonian dynamics of the qubit is exactly cancelled by the (conditional) measurement dynamics, so the qubit state is frozen at angle $\theta_s$.  The rate $E_c$ indicates how quickly the probability of this happening falls off as $|t_f-t_i|$ is increased.  Notably, the critical energy is a factor of 2 {\it smaller} than the Zeno (switching) rate derived in the main text.  We remind the reader that $r(t) = r_s$ does not mean just the single trajectory with that detector output, but encompasses the family of trajectories whose most likely value is $r_s$.  The physical paths that connect the states $|1\ra$ $(\theta=0)$ and $|2\ra$ $(\theta=\pi)$ (see Fig.~\ref{fig-phasespace}), have two finite action contributions to the probability that consist of bringing the state from $|1\ra$ to the vicinity of the fixed point, as well as taking the state from the fixed point to the state $|2\ra$.  This is only possible if $E > E_c$.  In the limit where the total time $T$ is much longer than the measurement time, $T \gg \tau$, the probability of staying near the fixed point will dominate the transitional pieces.   If $E < E_c$, the connecting trajectories will return the state back to its point of origin.

The fixed point \eqref{values} has been calculated exactly, with no smallness condition on $\Delta \tau$.  We note that since the stochastic energy is a measure of the likeliness of the solution (with ${\cal H} =0$ being most likely),  when $\Delta \tau$ is small compared to 1, such trajectories become more probable; this is also seen since $r_s \approx 1 + \Delta^2 \tau^2/2$ is only slightly larger than the ``measurement eigenvalue'' $r= + 1$, the average output of the detector when the system is in eigenstate $| 1\ra$.  

In the opposite limit, $\Delta \tau \gg 1$, the Hamiltonian dynamics will be the dominant effect, so it will be much more unlikely for the qubit state to get stuck at the singular point.  We see this from the fact that $E_c \rightarrow -\infty$, as well as $r_s \rightarrow \infty$ in this limit. More on this limit will be published later.


%

\end{document}